\documentstyle[12pt]{article}
\begin{document}
\def\complex{{\kern .1em {\raise .47ex \hbox {$\scriptscriptstyle
|$}}
\kern -.4em {\rm C}}}
\def\be {\begin{equation}}
\def\ee {\end{equation}}
\def\ba{\begin{array}}
\def\ea{\end{array}}
\def\ds{\displaystyle}
\def\tr{\mbox{Tr}}
\def\sqk{\sqrt{\kappa}\;}
\title{Classical and Quantum Intertwine}
\author{Ph. Blanchard \\
{\sl Faculty of Physics and BiBoS, University of Bielefeld}\\
{\sl D 33615 Bielefeld, FRG}
\and
A. Jadczyk \\
{\sl Institute of Theoretical Physics},
{\sl University of Wroc{\l }aw}, \\
{\sl Pl. Maxa Borna 9, PL 50 204 Wroc{\l }aw, Poland}}
\maketitle{}
\vspace{2cm}
\begin{abstract}{Model interactions between classical and quantum
systems are
briefly discussed. These include: general measurement-like couplings,
Stern-Gerlach experiment, model of a counter, quantum Zeno effect,
SQUID--tank
model.} \end{abstract}
%\newpage
\section{Introduction}
Heisenberg wrote in
his book \lq Physics and Philosophy\rq\, \cite{hei} that the Copenhagen
interpretation of Quantum Theory rests on a paradox, namely the
description of
quantum phenomena in terms of classical concepts. We know that Quantum
Mechanics
works extremely well  - it decribes and computes (although we would not
say: {\sl
explains}) not only those phenomena for which it was invented but also
numerous
others. But measurement-like processes cannot be described by
Schr\"odinger
equation. As emphasized by J. Bell \cite{bel}: \lq If, with
Schr\"odinger , we
reject extra variables, then we must allow that his equation is not
always
right\rq\, . Gisin and Percival formulated the same thesis as: \lq the
Schr\"odinger
equation is no longer the best for all practical purposes\rq\,
\cite{gis}.

In a recent paper \cite{bj} we propose a mathematically consistent model
of
interaction between classical and quantum systems, which provides an
answer to
the question of how and why quantum phenomena become real as a result of
interaction between quantum and classical domains. Our results show that
a simple dissipative time evolution can result in a dynamical exchange
of
information between classical and quantum levels of Nature. With a
properly
chosen initial state the quantum probabilities are exactly mirrored by
the state
of the classical system and moreover the state of the quantum subsystem
converges for $t\rightarrow +\infty$ to a limit which agrees with that
required by von Neumann-L\"uders standard quantum measurement projection
postulate. In our model the quantum system is coupled to a classical
recording device which will respond to its actual state. We thus give a
{\sl minimal mathematical semantics} to describe the measurement process
in
Quantum Mechanics. For this reason the toy model that we proposed can be
seen as the elementary building block used by Nature in the
communications
that take place between the quantum an classical levels. The model has
not
only nice mathematical properties but it is also of great practical
accesibility and, therefore, it is natural to formulate any practical
problem by starting from the general structure of the \lq Ansatz\rq\,
 we have
proposed.

What is the fundamental difference between a classical and a quantum
system?
A purely quantum mechanical description of a cat would need about
$10^{27}
$ independent parameters - and it would describe the cat as a closed
system -
thus certainly not the living cat. On the other hand to describe any
relevant property of the cat we have to specify far fewer parameters.
Having
first developed a full quantum theory of a cat and ignoring after that
all
the degrees of freedom that we are not able to measure or that come from
the
enviroment is to allow for a classical behavior. Insisting on necessity
of
\lq quantising\rq\, {\sl all} the degrees of freedom can be compared to
demanding
that only letters of an alphabet should be used - but not the
punctuation
marks! Sure, printed matter would then look impressively homogeneous.
But not sentence would ever end, no message would ever be transmitted.
The
Universe simply would not work that way.

In Section 2 we will briefly describe the mathematical and physical
ingredients of
the model. The purpose of Section 3 it to discuss the measurement
process
in this framework. The range of possible applications of the model is
rather wide as will be shown in Section 4 with a discussion of the
Zeno's
effect and in Section 5 with a description of the coupling between a
SQUID
and a damped classical oscillating circuit. Section 6 deals with some
concluding remarks.
\section{Interaction between a classical and a quantum system}
Let us briefly describe the mathematical framework we will use. A good
deal
more can be said and we refer the reader to \cite{bj,bjprep}.
Our aim is to describe a non trivial interaction between a quantum
system
$\Sigma_q$ in interaction with a classical system $\Sigma_c$. To the
quantum
system there corresponds a Hilbert space ${\cal H}_q$. In ${\cal H}_q$
we
consider a family of orthonormal projectors
$e_i=e_i^\star=e_i^2,\; (i=1,\ldots
,n) ,\; \sum_{i=1}^n e_i=1\, ,$ associated to an observable
$A=\sum_{i=1}^n
\lambda_i\, e_i .$ The classical system is supposed to have $m$ distinct
pure
states, and it is convenient to take  $m\ge n .$ The algebra ${\cal
A}_c$ of
classical observables is in this case ${\cal A}_c = \complex^m .$ The
set of
classical states coincides with the space of probability measures. Using
the
notation $X_c = \{s_0,\ldots ,s_{m-1})$, a classical state is an
$m-$tuple
$p=(p_0,\ldots ,p_{m-1}) ,\; p_\alpha\ge 0 ,\;
\sum_{\alpha=0}^{m-1}p_\alpha =1
.$ The state $s_0$ plays in some cases a distinguished role and can be
viewed as
the neutral initial state of a counter. The algebra of observables of
the total
system ${\cal A}_{tot}$ is given by
\be {\cal A}_{tot}={\cal A}_c \otimes L({\cal H}_q)
=\complex^m\otimes L({\cal H}_q)=\oplus_{\alpha =0}^{m-1}L({\cal H}_q) ,
\ee
and it is convenient to realize ${\cal A}_{tot}$ as an algebra of
operators on an
auxiliary Hilbert space ${\cal H}_{tot}={\cal H}_q\otimes
\complex^m=\oplus_{\alpha=0}^{m-1}{\cal H}_q .$ ${\cal A}_{tot}$ is then
isomorphic to the algebra of block diagonal $m\times m$ matrices
$A=diag(a_0,a_1,\ldots ,a_{m-1})$  with $a_\alpha\in  L({\cal H}_q).$
States on
${\cal A}_{tot}$ are represented by block diagonal matrices \be \rho =
diag
(\rho_0,\rho_1,\ldots ,\rho_{m-1})\ee where the  $\rho_\alpha $
 are positive trace class operators in
$L({\cal H}_q)$
with
$\sum_\alpha \tr(\rho_\alpha ) = 1 .$
By taking partial traces each state
$\rho $ projects on a \lq quantum state \rq
${\pi}_q(\rho )$
and a \lq classical state\rq
$\pi_c(\rho )$
given respectively by
\be \pi_q(\rho ) =\sum_\alpha \rho_\alpha,\ee
\be \pi_c(\rho ) =(\tr \rho_0,\, \tr \rho_1,\ldots ,\tr \rho_{m-1}).\ee

The time evolution of the total system is given by a semi group
$\alpha^t = e^{tL}$ of
positive maps \footnote{In fact, the maps we use happen to be also
{\sl completely}
positive.}
of ${\cal A}_{tot}$ -  preserving hermiticity, identity
and positivity - with $L$ of
the form \be L(A) = i[H , A ] +
{\sum}_{i = 1}^n \left ( V_i^{\star} A V_i - {1 \over 2} \{V_i^{\star}
V_i , A\}\right )
.\label{eq:lin} \ee
The $V_i$ can be arbitrary linear operators in $L({\cal H}_{tot})$ such
that $\sum V_i^\star V_i
\in {\cal A}_{tot}$ and $\sum V_i^\star A V_i \in {\cal A}_{tot}$
whenever $A\in {\cal A}_{tot}$,
$H$ is an arbitrary block-diagonal self adjoint operator
$H=diag(H_\alpha )$ in ${\cal H}_{tot}$
and $\{\; ,\;\}$ denotes anticommutator i.e.
\be \{A\; , B\;\} \equiv AB+BA .\ee
In order to couple the given quantum observable $A=\sum_{i=1}^n
\lambda_i\, e_i $ to the classical
system, the $V_i$ are chosen as tensor products $V_i=\sqk
e_i\otimes\phi_i$, where $\phi_i$ act as transformations on classical
(pure)
states. Denoting  $\rho (t) = \alpha_t (\rho(0))$, the time evolution of
the
states is given by the dual Liouville equation  \be {\dot \rho}(t) =
-i[H ,\rho
(t)] + {\sum}_{i = 1}^n \left ( V_i \rho (t) V_i^{\star} - {1 \over 2}
\{V_i^{\star} V_i , \rho (t)\} \right ), \label{eq:lio}
\ee
where in  general $H$ and the $V_i$ can explicitly depend on time.\hfill
\\
{\bf Remarks:}\hfill
\\
1) It is possible to generalize this framework for the case where the
observable $A$ that is being measured admits a continuous spectrum (as
for
instance in a measurement of the position) with $A={\int_{\bf R}
\lambda\,
dE(\lambda )}.$ See \cite{bjprep,bjsq} for more details and Sections 3.3 and 5
where
concrete examples of a situation of this type will be briefly
described.\hfill\\
2) Since the center of the total algebra ${\cal A}_{tot}$ is invariant
under
any automorphic unitary time evolution, the Hamiltonian part $H$ of the
Liouville operator is not directly involved in the process of transfer
of
information from the quantum subsystem to the classical one. Only the
dissipative part can achieve such a transfer in a finite time.
\section{The measurement process and all that}
In \cite{bj} we propose a simple, purely dissipative Liouville operator
(i.e. we
put $H=0$) that describes an interaction of $\Sigma_q$ and $\Sigma_s$,
for
which $m=n+1$ and $V_i=e_i\otimes\phi_i,$ where $\phi_i$ is the flip
transformation of $X_c$ transposing the neutral state $s_0$ with $s_i$.
We show that the Liouville equation can be solved explicitly for any
initial state $\rho(0)$ of the total system. Assume now that we are able
to prepare at time $t=0$ the initial state of the total system
$\Sigma_{tot}$ as an uncorrelated product state
$\rho (0) = w\otimes P^{\epsilon }(0),$
$P^{\epsilon }(0) = (p_0^{\epsilon },p_1^{\epsilon },\ldots
,p_n^{\epsilon })$ as initial state of the classical system parametrized
by $\epsilon$, $0\leq \epsilon\leq 1 :$
\be
p_0^{\epsilon}=1-{n\epsilon\over n+1},
\ee
\be
p_i^{\epsilon}={\epsilon\over n+1} .
\ee
In other words for $\epsilon = 0$ the classical system starts from the
pure state $P(0) = (1,0,\ldots ,0)$ while for $\epsilon =1 $ it starts
from the state $P^{'}(0) = ({1\over n+1},{1\over n+1},\ldots ,{1\over
n+1})$ of maximal entropy. Computing $p_i(t) = \tr (\rho_i(t))$ and
then the normalized distribution
\be {\tilde p }_i(t) = {p_i(t) \over {{\sum}_r p_r(t) }}\ee
with $\rho(t)=(\rho_0(t),\rho_1(t),\ldots ,\rho_n(t))$ the state of the
total system we get:
\be
{\tilde p }_i(t) = q_i + { {\epsilon (1-nq_i)}  \over {\epsilon n+{{
(1-\epsilon )(n+1)}
\over 2} (1-e^{-2t}) }} ,\label{eq:pti}\ee
where we introduced the notation \be q_i = \tr (e_i w) ,\ee for the
initial quantum probabilities to be measured.
For $\epsilon =0$ we have
${\tilde p }_i(t)=q_i$ for all $t > 0,$ which means  that the quantum
probabilities are exactly, and immediately after switching on of the
interaction,
mirrored by the state of the classical system. For $\epsilon = 1$ we get
${\tilde
p }_i(t)=1/n.$ The maximum entropy state is a stationary state of
$\Sigma_c$ and
in this case we get no information at all about the quantum state by
recording the
time evolution of the classical one. For $\epsilon =0 , $ that is when
the
measurement is exact, we get for the partial quantum state
$$\pi_q(\rho (t)) = \sum_i e_i w
e_i+e^{-t}(w-\sum_i e_i w e_i) ,$$
 so that
\be
\pi_q(\rho (\infty )) =\sum_i e_i w e_i ,
\ee
which means that the partial state of the quantum subsystem $\pi_q(\rho
(t))$
tends for $t\rightarrow +\infty$ to a limit which coincides with the
standard von
Neumann-L\"uders quantum measurement projection postulate.\\
\\
{\sl 3.1. Efficiency versus accuracy by measurement}\\
\\
Let us consider the case where
\be
V_i=\sqk e_i\otimes f_i ,
\ee
$f_i$ being the transformation of $X_c$ mapping $s_0$ into $s_i$. In the
Liouville equation we consider also an Hamiltonian part. The Liouville
equation implies:
\be
{\dot \rho_0}=-i\left[ H,\rho_0\right] - \kappa \rho_0 ,
\ee
\be
{\dot \rho_i}=-i\left[ H, \rho_i\right] + \kappa \rho_0 e_i ,
\ee
where we allow for time dependence i.e. $H=H(t) ,$
$e_i=e_i(t) .$ Setting $r_0(t)=\tr (\rho_0 (t)) ,$  $r_i(t)=\tr
(\rho_i(t)),$ and
assuming that the initial state is of the form $\rho=(\rho_0,0,\ldots
,0)$ we
conclude that ${\dot r_0}=-kr_0$ and thus $r_0(t)=e^{-kt}$ which implies
that
for small $t$
\be
\sum_{i=1}^n r_i(t)=1-e^{-kt}\approx kt ,
\ee
from which it follows that a $50\%$ efficiency requires $\log 2 /\kappa$
time of
recording. It is easy to compute $r_i(t)$ and $${\tilde p}_i(t)=
{r_i(t)\over {\sum_{j=1}^n r_j(t)}}$$ for small $t$. One obtains
\be
{\tilde p}_i(t)=q_i+{\kappa^2t^2\over 2}{1\over \kappa}\langle{de_i\over
dt}\rangle_{\rho_0} ,
\ee
where
\be
{de_i\over dt}\doteq {\partial e_i\over {\partial t}}+i\left[
H,e_i\right] .
\ee
Efficiency requires $\kappa t\gg 1$ while accuracy is achieved if
$(\kappa
t)^2\ll {\kappa\over {\langle{\dot e}_i\rangle_{\rho_0}}} .$ To
monitor effectively {\sl and}
accurately fast processes we must therefore take at least $\kappa\approx
10^2
{\langle{\dot e}_i\rangle_{\rho_0}} .$ Suppose now that $H$ and $e_i$ do
not
depend on time. Then it is easy to show that if either $\rho_0(0)$ or
$e_i$
commutes with $H ,$ we get ${\tilde p}_i(t)=q_i$ exactly and instantly.\\
\\
{\sl 3.2. Stern Gerlach experiment}\\
\\
In the spirit of A. B\"ohm (cf. Ref. \cite[ Ch. XIII]{ab})
 we model a Stern--Gerlach device by a pure
spin
$1/2$ particle interacting with a spinless atom. Assuming that the
magnetic
field is linear in $z$ the interaction Hamiltonian can be written
\be
H_{int}=2\mu_B B z \sigma_3 .
\ee
Writing
$$
\sigma_3={\vert\uparrow\rangle\langle\uparrow\vert-
\vert\downarrow\rangle\langle\downarrow\vert \over 2},
$$
$H_{int}$ is now given by
\be H_{int}=\mu_B B \left(\vert\uparrow\rangle\langle\uparrow\vert\;z-
\vert\downarrow\rangle\langle\downarrow\vert\;(-z)\right) .
\ee
Supposing now that the atom can be directly observed we can replace it
for all
practical purposes by a $3-$state classical device $(s_0,s_+,s_-).$ The
coupling
is then modelled by
\be
\sqk \left(p\; {\sl flip} (0\rightarrow +) + (1-p)\;{\sl flip}
(0\rightarrow -
)\right)
\ee
and we are now in position to approximate Stern--Gerlach experiment by
our
$3-$state model. For more details see also \cite{bjprep}.\\
\\
{\sl 3.3 Model of a counter}\\
\\
We consider a one-dimensional quantum mechanical particle. The counter
sensitivity is described by an operator valued function $f(t)$ and the
quantum
system $\Sigma_q$ with ${\cal H}_q=L^2({\bf R},dx)$ is coupled to a
$2-$state
classical system. The Liouville equation for the state of the total
system is
\be
{\dot \rho} = - i \left[ H,\rho\right] + V\rho V^{\star} -{1\over
2}\left\{
V^{\star} V, \rho\right\} ,
\ee
with
\be
V=f\otimes
\pmatrix{0&1\cr 0&0\cr} =\pmatrix{0&f\cr 0&0\cr}
\ee
Now explicitly we obtain
\be
{\dot\rho}_0=-i\left[ H,\rho_0\right]-{1\over
2}\left\{f^{\star}f,\rho_0\right\},
\ee
\be
{\dot\rho}_1=-i\left[H,\rho_1\right]+f^{\star}\rho_0 f .
\ee
Taking $H={1\over i}{d\over dx}$ and $f=f^{\star}=f(x,t)$ we obtain for
the
counting rate ${\dot p}_1(t)$ in a free evolving state:
\be
{\dot p}_1(t)={\int_{\bf R}\vert \Psi (x-t)\vert^2 f^2(x,t)e^{-{\int_0^t
f^2(x+s-t,s)\,ds}}\,dx}.
\ee
Assume now that we have to do with a point particle i.e.
$$\vert\Psi (x)\vert^2=\delta (x-x_0)$$
we obtain in this idealized case
\be
{\dot p}_1(t)=f^2(x_0+t,t)e^{-{\int_0^t f^2(x_0+s,s)\,ds}}
\ee
which expresses the fact that the counting rate depends on how long the
"detector" was already in contact with the particle. Let us remark that
the
quantum particle we consider is as in \cite{pas} an ultra--relativistic
one. For
more details see \cite{bjprep}.
\section{Quantum Zeno's Effect}
The standard view of evolution of quantum states can be described as
follows.
Quantum states evolve trough establishment of coherent superpositions.
An initial
state $\Psi_0$ which is unstable develops into a superposition
$\Psi_t=a_0\Psi_0+a_d\Psi_d$ of undecayed and decay-product states. A
measurement of the survival probability
$\vert\langle\Psi_t,\Psi_0\rangle\vert^2$
projects the state $\Psi_t$ back to the initial undecayed state;
repeated,
frequent measurements can inhibit or even prevent the decay. This is
called the
Quantum Zeno effect. W. Yourgrau \cite{you} attributed it to
A.M. Turing
while
B. Misra and E.C.G. Sudarshan \cite{ms} coined the name and discussed the
relevant timescales at work. Using our model of a continuous measurement
we can
easily discuss this effect for a quantum spin $1/2$ system coupled to a
$2$-state
classical system \cite{zen}. We consider only one orthogonal projector
$e$ on the
Hilbert space ${\cal H}_q=\complex^2 .$ To specify the dynamics we
choose the
coupling operator $V$ in the following symmetric way:
\be
X=\sqk \pmatrix{0&e\cr e&0\cr}.
\ee
The Liouville equation for the total state $\rho=diag(\rho_0,\rho_1)$
reads now:
\be
{\dot\rho_0}=-i\left[H,\rho_0\right]+\kappa (e\rho_1e-{1\over
2}\{e,\rho_0\}),
\ee
\be
{\dot\rho_1}=-i\left[H,\rho_1\right]+\kappa (e\rho_0 e-{1\over
2}\{e,\rho_1\}).
\ee
The partial quantum state $\pi_q(\rho )={\hat \rho}=\rho_0(t)+\rho_1(t)$
evolves
in this particular model independently of the state of the classical
system,
which expresses the fact that we have here only transport of information
from
$\Sigma_q$ to $\Sigma_c .$ The time evolution of ${\hat \rho} (t)$ is
given by
\be
{\dot{\hat\rho}}=-i\left[H,{\hat\rho}\right]+\kappa (e{\hat\rho}e-
{1\over 2}\{e,{\hat\rho}\})\label{eq:zew}
\ee
Let us now choose the Hamiltonian part $H={\omega\over 2}\sigma_3$ and
$e={1\over 2}(\sigma_0+\sigma_1)$, and to start with the quantum system
$\Sigma_q$
being for $t=0$ in the eigenstate of $\sigma_1.$ We repeatedly check
with
frequency $\kappa$ if the system is still in this initial state, each
\lq
yes\rq\, inducing a flip in the coupled classical device, which we
continuously
observe. The solution of (\ref{eq:zew}) such that ${\hat\rho}(0)=e$ can be
easily
found. Moreover it is possible for strongly coupled system i.e. for
$\kappa
 t\gg1$ and $\kappa /\omega\gg 1$ to obtain asymptotic formulae for the
distance
travelled by the quantum state $d({\hat\rho}(t),e)$ in the Bures or in
the
Frobenius norm $\Vert{\hat\rho}\Vert^2 =\tr ({\hat\rho}^2) .$ In this
asymptotic
regime we can show that the Bures distance achieved during the coupling is
given by
\be
d({\hat\rho}(t),e)\approx \omega\sqrt{{t/ \kappa}}.
\ee
The effect of slowing down the evolution of the quantum system can be
confirmed
by an independent,strong but non-demolishing, coupling of a third
classical
device.  In \cite{sto} we show moreover that a piecewise deterministic
Markov
process taking values on pure states of the total system is naturally
associated
to the Liouville equation. \hfill
\\
\\
{\sl 4.1. Comments on \lq meaning of the wave function\rq\,}\hfill
\\
\\
One could think of using the Quantum Zeno effect for slowing down the time
evolution, so that the state of a quantum system can be determined by
measurements of sufficiently many observables. This idea, however,
would not work, similarly like would not work
the idea of \lq protective measurements\rq\,
of Y. Aharonov et. al. (cf. Refs \cite{av1,av2}). To apply Zeno--type
measurement, similarly like to apply a \lq protective measurement\rq\, one
would have to know the state beforhand. This negative statement
 does not mean that the quantum
state cannot be determined. To the contrary, by coupling a quantum
system to a
3--state asymmetric (but {\sl not} symmmetric!) device according to our
recipe (cf. \cite{bjprep} for more details),
and using sufficiently strong and sufficiently short couplings,
one can measure an arbitrary
finite number of expectation values of quantum observables,
without disturbing
the quantum system beyond prescribed limits.
Far from being able to determine
the quantum state with an {\sl infinite precision},
such a, theoretically feasible, sequence of measurements
should be sufficient for all practical purposes.\footnote{
For some quantum system there will be, of course, unsurmountable
 technical
difficulties} Moreover, it is natural to try to use the
progressive,
incomplete,
 knowledge of the quantum state for slowing down the evolution
during next series of measurements. This last idea
needs however further investigations.
\section{SQUID coupled to a damped classical
oscillator}
In a superconducting ring of uniform thickness the quantized flux
states do not interact. The quantum state of the ring can only be
changed by
warming it up, changing the applied external field and then cooling it
down to its
superconducting temperature again. However, mixing of the flux states
becomes possible if the ring has a so-called weak link, across which the
magnetic
flux can leak. These objects are called SQUIDS and posses a wide variety
of
macroscopic quantum mechanical properties. In recent years there has
been
considerable discussion of the dynamics of a model quantum--mechanical
system
consisting of a SQUID coupled to a dissipative classical linear
oscillator
\cite{spi1,spi2,ral1,ral2}. Our aim is to briefly show that our
framework is very
well adapted to give a description of a \lq mini\rq SQUID and to discuss
the
behaviour of the coupled system consisting of a macroscopic classical
system
(tank circuit) and a single quantum object (SQUID). The states of the
total
system are now described by $\rho_t (\phi ,\pi )$, with $\rho_t(\phi
,\pi )\ge 0$
and $$
{\int_{{\bf R}^2} \tr (\rho_t (\phi ,\pi ))\, d\phi\, d\pi} = 1 .
$$
The Liouville equation that describes SQUID+tank circuit is:
\be
\ba{ll}
{\partial\rho (\phi,\pi,t) \over {\partial t}}=&
-{i\over \hbar}\left[ H(\phi ),\rho(\phi ,\pi
,t)\right]-\pi\,{\partial\rho \over {\partial t}}+\cr\cr
& +\left({\pi\over RC}+{\phi\over LC} -
{1\over C}I_{IN}(t)\right){\partial\rho \over {\partial\pi}}+\cr\cr
& +{\rho\over RC}+(L\rho )(\phi ,\pi ,t),
\ea  \ee
where
\be
\ba{ll}
L\rho (\phi, \pi, t)=&\alpha{\int_{\bf R} f(\Phi-\Phi_{EXT}-a)\rho(\phi ,
\pi
-a,t)f(\Phi-\Phi_{EXT}-a) \, da}\cr\cr
& -\alpha\beta\rho (\phi , \pi ,t),
\ea
\ee
$f(x)=f(-x)\ge 0$ being a real function characterising the coupling
(e.g. a
Gaussian), $\beta={\int f(x)^2\,dx}$, and $\alpha\beta=\mu /
(C\Lambda)$, with
$R,C,L$ characterizing electric properties of the tank circuit, $\mu$
the mutual
SQUID--tank inductance  while $\Lambda$ the superconducting ring
geometry.
The SQUID Hamiltonian is defined on the Hilbert space ${\cal
H}_q=L^2\left({\bf
R},d\Phi\right)$
\be
H(\phi)={Q\over 2C}+{(\Phi - \Phi_{EXT} )\over
{2\Lambda}}-\hbar\omega\cos\left
({\Phi\over\Phi_0}\right),
\ee
\be
\Phi_{EXT}=\phi_{ext}+\mu\phi,
\ee
\be
Q=-i\hbar{d\over{d\Phi}},
\ee
\be
\Phi_0={\hbar\over{2e}}.
\ee
Notice that the SQUID Hamiltonian depends explicitly on the classical
parameter
$\phi$. It can be now shown (see \cite{bjprep}) that the expectation
value of
$\phi$ satisfies the non-linear equation postulated and investigated in
Refs. \cite{ral1,ral2}. The same method that we used here for
modelling of SQUID-tank coupling can be applied as well to other
problems where the classical system is expected to respond to
\lq averages\rq\, of some quantum observables.
\footnote{Like, for instance, classical gravitational field
is expected to have as its source averaged energy--momentum
tensor of quantized matter.}
\section{Concluding remarks}
The exceptionally brilliant calculational successes of Quantum Mechanics
cannot
cause to forget the degree of conceptual confusion still present. The
essential
problem follows from the fact that Quantum Mechanics is the most
fundamental
theory we know. But if it is really fundamental, it should be
universally
applicable. In particular quantum physics should be able to explain also
the
properties of macroscopic objects and occurence of macroscopic events.
But
measurement situations show clearly that it is impossible to apply
standard
Quantum Mechanics in a consistent way to all relevant situations. If
there is
such a universal theory, it is therefore not Quantum Theory.\hfill
\\
Indeterminism is an implicit part of classical physics. Irreversible
laws are
fundamental and reversibility is an approximation. We cannot refrain
quoting from R. Haag's paper {\sl Irreversibility introduced on a fundamental
level} \cite{haa}: \lq  ... once one accepts in\-de\-ter\-mi\-nism,
there is no reason
against including irreversibility as part of the fundamental laws of Nature.
\rq\. We propose to consider
$\Sigma_{tot}=\Sigma_q\otimes\Sigma_c$ and the behaviour associated to
the total
algebra of observables ${\cal A}_{tot}=\otimes{\cal A}_q\otimes{\cal
A}_c
=C(X_c)\otimes L({\cal H}_q)$ is now taken as the fundamental reality
with
{\sl pure} quantum behaviour as an approximation valid in the
exceptional cases
when dissipative effects can be neglected. In ${\cal A}_{tot}$ we can
describe
irreversible changes occuring in the physical world -- like the
formation of a
track in bubble chamber or the tragical cat's death -- as well as
(idealized)
reversible pure quantum processes.

Lot of work must still be done, lot
of
prejudices overcomed. What we propose does not aspire to be a magic
medicine
that will rejuvenate Quantum Theory and make it
Universal-And-True-For-Ever.
But, perhaps, it will help to stop the bleeding from some open scars..

\end{document}